\documentclass[showpacs, twocolumn, prd]{revtex4}

\usepackage{amssymb, amsmath} 

\usepackage{bbm} 

\usepackage[english]{babel}
\usepackage{slashed}

\usepackage{graphicx}
\usepackage{color}

\usepackage[bookmarks,breaklinks,plainpages=false,pdfpagelabels]{hyperref}  
	\hypersetup{linkcolor=red,citecolor=blue,filecolor=dullmagenta,urlcolor=darkblue} 

\definecolor{grey}{rgb}{0.4,0.4,0.4}
\definecolor{dullmagenta}{rgb}{0.4,0,0.4}
\definecolor{darkblue}{rgb}{0,0,0.4}
\definecolor{orange}{rgb}{1,0.5,0}
\definecolor{lightbrown}{rgb}{0.75,0.5,0.25}
\definecolor{tan}{cmyk}{0.14,0.42,0.56,0}
\definecolor{djunglegreen}{cmyk}{0.99,0,0.52,0}
\definecolor{lightgreen}{rgb}{0,1,0}
\definecolor{olivegreen}{cmyk}{0.64,0,0.95,0.40}
\definecolor{midgreen}{rgb}{0.0,0.675,0.0}

\renewcommand{\l}{\ensuremath{\lambda}}
\newcommand{\s}{\ensuremath{\sigma}}

\newcommand{\q}{\quad}

\newcommand{\vs}{\vspace}

\renewcommand{\.}{\hspace{0.5mm}}

\newcommand{\ra}{\ensuremath{\rightarrow}}

\newcommand{\Drm}{\ensuremath{\mathrm{D}}}

\newcommand{\Vrm}{\ensuremath{\mathrm{V}}}

\newcommand{\irm}{\ensuremath{\mathrm{i}}}

\newcommand{\Ocal}{\ensuremath{\mathcal{O}}}

\newcommand{\Scal}{\ensuremath{\mathcal{S}}}

\newcommand{\Ibbm}{\ensuremath{\mathbbm{I}}}

\renewcommand{\d}{\ensuremath{\mathrm{d}}}
\newcommand{\ee}{\ensuremath{\mathrm{e}}}

\newcommand{\defas}{\mathrel{\mathop :}=} 
\newcommand{\hph}[1]{\hphantom{#1\;\,}}

\newcommand{\eg}{e.g.}
\newcommand{\ie}{i.e.}

\newcommand{\hc}{{\rm H.c.}}

\newcommand{\cf}{c.f.}

\makeatletter
  \def\ifundefined#1{\expandafter\ifx\csname#1\endcsname\relax}
  \ifundefined{default@color}
    \let\default@color=\current@color
  \fi
\makeatother

\newcommand{\beq}{\begin{equation}}
\newcommand{\eeq}{\end{equation}}
\newcommand{\bea}{\begin{eqnarray}}
\newcommand{\beas}{\begin{eqnarray*}}

\newcommand{\eea}{\end{eqnarray}}
\newcommand{\eeas}{\end{eqnarray*}}

\begin{document}

\title{Axion arising from warped extra-dimensional gauge fields}

\author{Y.~Burnier}
\affiliation{Department of Physics and Astronomy, SUNY,\\
Stony Brook, New York 11794-3800, USA}
\email{yburnier@notes.cc.sunysb.edu}

\author{F.~K{\"u}hnel}
\affiliation{Arnold Sommerfeld Center, Ludwig-Maximilians University,\\
Theresienstr.~37, 80333 M{\"u}nchen, Germany}
\email{florian.kuehnel@physik.lmu.de}

\date{\today}

\begin{abstract}
We present a connection between two known solutions to the strong-{\sc cp} problem: the standard introduction of axions and the extra-dimensional one, relying on topological arguments. Using an equivalent lower-dimensional set-up with a warped extra-dimension but without adding any new fields, it is shown that an additional light degree of freedom appears. Like an axion, it couples to the topological charge density via Fermionic loop corrections. Its decay constant is related to the geometry of the extra dimension and is suppressed by the warping scale.\\
\end{abstract}


\maketitle

\section{Introduction}
\label{sec:Introduction}

Quantum chromodynamics has a complicated vacuum structure \cite{Jackiw:1976pf}. Due to the nontrivial group of mappings from the coordinate space to the gauge group, there is an infinite number of vacua parametrized by an angle $\theta$. The physics depends on $\theta$, which becomes a new parameter of the theory. Equivalently this may be expressed by a new {\sc cp}-violating effective term in the Lagrangian of the form $\theta\mspace{1mu}q$, where $q$ is the topological charge density \footnote{In (3+1)-dimensional {\sc qcd} $q = \frac{ g^{2} }{ 32\mspace{1mu}\pi^{2^{^{^{}}}}\! }\.F^{\mu\nu} \widetilde{F}_{\mu\nu.}$, while in (1+1)-dimensional {\sc qed} $q = \frac{ e }{4\mspace{1mu}\pi}\.\varepsilon_{\mu\nu}\mspace{1mu}F^{\mu\nu}$.$^{^{^{}}}_{_{_{_{_{}}}}}$}.

The presence of massive Fermions also leads to an effective topological term proportional to the imaginary part of the Fermionic mass matrix determinant $\det \mathcal{M}$. The fact that the sum $\theta + \arg \det \mathcal{M}$ is experimentally constrained to be smaller than $2 \times 10^{- 10}$ is known as the strong-{\sc cp} problem. Note that chiral symmetry can be used to redefine the Fermion masses and get rid of their imaginary part. However because of the chiral anomaly, this amounts to shifting the value of $\theta$ so that the sum $\theta + \arg \det \mathcal{M}$ remains constant.

Extra-dimensional solutions to the strong-{\sc cp} problem were proposed some time ago \cite{Khlebnikov:1987zg, Chaichian:2001nx, Khlebnikov:2004am}. The formal idea in Ref.~\cite{Khlebnikov:2004am} is that the presence of the anomaly depends on the number of dimensions and vanishes, for instance, if one extra-dimension is added. If the Fermionic anomaly actually vanishes, one is free to rotate away the imaginary phase of the Fermion masses, which suggest that the strong-{\sc cp} problem will disappear automatically. Of course the solution is not that simple. One would like that after integrating out the extra-dimension, the low-energy effective theory resembles the normal {\sc qcd}, so that one has to show that the strong-{\sc cp} problem does not reappear after proper localization of the fields \cite{Bezrukov:2008da}.

The models considered in Ref.~\cite{Khlebnikov:2004am, Bezrukov:2008da} are not very realistic from the cosmological point of view. In these models, the spatial dimensions have the topology of a sphere and the fields are localized on the equator. The reason being the observation of Ref.~\cite{Khlebnikov:2004am} that the Fermionic anomaly reappears if boundaries are present, preventing the resolution of the strong-{\sc cp} problem in orbifold models. Indeed, if the spatial dimensions take the form of $\mathbb{R}^{d} \times \Ibbm$, where $\Ibbm$ is the interval $[ - L , L ]$, the anomaly in the Fermionic current $J^{A}$ was shown to be \cite{ArkaniHamed:2001is}\\[-5mm]
\begin{align}
	\partial_{A} J^{A}( x^{\mu}, z )
		&=						q( x^{\mu}, z ) \big[ \delta( z - L ) + \delta( z + L ) \big]
								\, .
								\label{eq:anom}
								\\[-6mm]
								\notag
\end{align}

We want to have an alternative look at this restriction. The strong-{\sc cp} problem might actually still be solved if specific boundary conditions are imposed on the gauge fields, namely if the topological charge vanishes at the boundaries:
\vs{-2.5mm}
\begin{align}
	q \big|_{z\.=\.\pm L}
		&=						0
								\; .
								\label{eq:BC0}
								\\[-6mm]
								\notag
\end{align}
If the strong-{\sc cp} problem should be solved by this condition, it is not clear how. In the models of Ref.~\cite{Khlebnikov:2004am, Bezrukov:2008da}, the spherical topology changes the group of mappings from the coordinate space to the gauge group, so that the degeneracy of the vacua is removed. In the orbifold case, it is not so as the space manifold factorizes. The resolution of this paradox is easy to guess: If the strong-{\sc cp} problem is not solved by the topology, it has to be solved by the appearance of an axion-like particle \cite{Peccei:1977hh}.

Different extra-dimensional models with axions were already considered in the literature \cite{Dienes:1999gw}. However, they either required the addition of another field or extra terms to the Lagrangian to get the particular coupling of the axion to the topological charge.

In our set-up, the boundary conditions are enough to solve the strong-{\sc cp} problem and no other fields need to be introduced. As a first step to simplify the problem, we study a topologically-equivalent lower-dimensional model with the same vacuum structure: The two-dimensional Abelian Higgs model. This model has the same amount of vacuum degeneracy as four-dimensional {\sc qcd} and was already studied many times as a toy model for {\sc qcd}, \eg, the $U( 1 )$ problem was first solved there \cite{Kogut:1974kt}. This solution could then be mapped to {\sc qcd} \cite{tHooft}.

The resolution can be sketched as follows: In our model, one single field can provide an additional scalar, namely the extra-dimensional component of the gauge field $A_{2}$. This component actually couples to Fermions with the matrix $\gamma^{}_{5}$, which is just what we need to get a topological effective coupling to the photon $A_{\mu}$ via Fermionic loop corrections from the diagram
\vs{-2mm}
\begin{align}
	\put(0,-16){\includegraphics[scale=1]{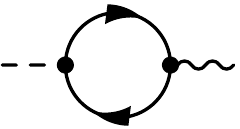}}
		&\mspace{131mu}\to\;		A^{}_{2}\.\varepsilon_{\mu\nu}\.F^{\mu\nu}
								\; .
								\label{eq:qcor}
								\\[-7mm]
		&						\notag
\end{align}

In the following, we will show explicitly that the different parameters of the model can be chosen to make it realistic.

\section{Abelian Higgs model}
\label{sec:Abelian-Higgs-model}

The action for the Abelian Higgs model in 1+1 dimensions reads \footnote{The signature of the metric is $( +, - )$, and Greek indices run over $0, 1$.}
\begin{align}
\begin{split}
	\Scal
		&=						\int\! \d^{2} x \left[
								-\, \frac{1}{4}\.F^{}_{\mu \nu }F_{}^{\mu \nu}
								- \Vrm( H )
								+ \frac{1}{2}\left| \Drm_{\mu} H \right|^{2}
								\right.
								\\
		&\hph{=}					\phantom{\int\! \d^{2}_{} x\mspace{8mu}}
								\left.
								\vphantom{\frac{ 1 }{ 2 }}
								+ \irm\.\bar{\Psi}\.\gamma^{\mu} \Drm_{\mu}\Psi
								- \tilde{\mu}\.\Psi^{\dagger}_{\sf L}\.\Psi^{}_{\sf R}
								- \tilde{\mu}^{*}\.\Psi^{\dagger}_{\sf R}\.\Psi^{}_{\sf L}
								\right]
								,
								\label{eq:Lmod1}
\end{split}
\end{align}
with the covariant derivative $\Drm_{\mu} \defas \partial _{\mu} - \irm\.\tilde{ e }\.A_{\mu}$. Here and below we use a tilde to denote the (1+1)-dimensional variables.

As emphasized in the introduction, the Abelian Higgs model has the same complicated vacuum structure as {\sc qcd}. In mathematical terms this is described by the fact that the specific homotopy groups match: $\pi^{3}\big( SU( 3 ) \big) = \pi^{1}\big( U( 1 ) \big) = \mathbb{Z}$. The Abelian Higgs model also has a strong-{\sc cp}-like problem, namely the physics depends on the $\theta$-vacua by the effective term $\theta\.\frac{ e }{4\mspace{1mu}\pi}\.\varepsilon_{\mu\nu}\mspace{1mu}F^{\mu\nu}$. 

The Higgs has a marginal r{\^o}le here, it is only needed to add an extra component to the gauge field; (1+1)-dimensional massless electromagnetism has no degrees of freedom, no photon and is therefore a poor toy model for {\sc qcd}. Adding the Higgs, renders the photon massive and gives a longitudinal mode{\.---\.}a toy particle for the {\sc qcd} gluon. For the comparison to {\sc qcd} one should consider the limit where the photon mass becomes much smaller than any other scale in the model.

\section{Extra-dimensional model}
\label{sec:Extra--dimensional-model}

We will now define a (2+1)-dimensional Abelian Higgs model, where the additional dimension is the interval $\Ibbm = [ - L , L ]$ and the fields are localized on a brane at the middle of the interval. The low-energy theory will match the previous Abelian Higgs model with an additional axion particle.

{\bf Bosonic sector:} Here we have \footnote{Capital Latin indices are out of $\{ 0, 1, 2 \}$, and the coordinates are denoted by $x^{0} = t$, $x^{1} = x$ and $x^{2} = z$.}
\begin{align}
	\Scal
		&=						\int \d^{2} x\; \d z\; \Delta
								\left[
								-\, \frac{ \left( F_{M N} \right)^{2} }{ 4 }
								+ \frac{ \left| \Drm_{M} H \right|^{2} }{ 2 }
								- \Vrm( H )
								\right]\mspace{-0.5mu}
								,
								\label{eq:LH}
\end{align}
with the Higgs potential $\Vrm( H ) = \frac{\lambda}{4}\mspace{-1mu} \left( | H |^{2} - v^{2} \right)^{2}$, $\l > 0$, $v > 0$, and $\Drm_{M} \defas \partial_{M} - \irm\mspace{1.5mu}e\mspace{1mu}A_{M}$. The Higgs field can be decomposed in real fields as $H = ( h + v )\.\ee^{\irm\.\sigma}$, $h$ has mass $m_{H}^{2} = 2\.\lambda\.v^{2}$ and the gauge field gets a mass $m_{W} = e\.v$.

Note that, in this work, we will assume that the `warp factor' $\Delta( z )$, which is used to localize the fields \cite{nonAlocal, warpedgf}, is coming from the coupling to some external classical field. This is different from gravity, and done for convenience.

{\bf Fermionic sector:} Fermions are localized on the brane with a domain-wall-like mass functions $m_{j}(z)$. Two Fermions are needed in 2+1 dimensions to get one Dirac Fermion in 1+1 dimensions. We add to action \eqref{eq:LH} the part \footnote{We use the convention, wherein the gamma matrices take the form $\gamma_{0} = \s_{1}$, $\gamma_{1} = \irm\.\s_{2}$, $\gamma_{2} = \irm\.\s_{3}$, and thus $\gamma_{5} = \irm\.\gamma_{2}$.}
\vs{-1mm}
\begin{align}
								\int \d^{2} x\; \d z\;
								\Big[\.
								\bar{\Psi}_{j}
								\!\left(
								\.\irm\.\slashed{\Drm}
								+ m_{j}
								\right)\!
								\Psi_{j}
								- \left(
								\mu\.\bar{\Psi}^{}_{1} \Psi^{}_{2}
								+ \hc
								\right)\!
								\mspace{1mu}\Big]
								\mspace{3mu} ,
								\label{eq:L-Fermion-3D}
\end{align}
wherein the sum over $j = 1, 2$ is implicit.

The choice of the boundary conditions at the end points of the extra dimension have to fulfil two criteria: The vanishing of the anomaly \eqref{eq:anom} and the conservation of energy. The latter implies that the component $T_{0 2}$ of the energy-momentum tensor vanishes at $z = \pm L$. Both conditions are satisfied if and only if we impose \footnote{Note that the boundary condition \eqref{eq:F01-vanishes-on-the-boundary} is not the usual one, chosen to cancel the boundary terms arising when varying $A_{M} \ra A_{M} + \delta A_{M}$ and integrating the corresponding term in the action by parts. Hence, one has to introduce additional boundary terms in the equation of motion, however they play no r{\^o}le in our particular setup.}
\begin{subequations}
\begin{align}
	\partial_{2} \sigma \big|_{z\.=\.\pm L}
		&=						e\.A_{2} \big|_{z\.=\.\pm L}
								\; ,
								\label{eq:partial-two-sigma=a2}
								\displaybreak[1]
								\\[1mm]
	\partial_{2} h\.\big|_{z\.=\.\pm L}
		&=						0
								\; ,
								\label{eq:boundary-condition-for-h}
								\displaybreak[1]
								\\[1mm]
	F^{}_{0 1} \big|_{z\.=\.\pm L}
		&=						0
								\; ,
								\label{eq:F01-vanishes-on-the-boundary}
								\displaybreak[1]
								\\[1mm]
	\Psi^{\dagger}_{j}\.\Drm^{}_{2} \Psi^{}_{j} \big|_{z = \pm L}
		&=						0
								\; .
								\label{eq:Boundary-condition-Fermion}
\end{align}
\end{subequations}
The remaining task is to check that these boundary conditions indeed lead to the desired low-energy theory on the brane. To achieve that, we have at hand the warp factor $\Delta( z )$ and the domain-wall functions $m_{j}( z )$.

\section{Dimensional Reduction}
\label{sec:Dimensional-Reduction}

We perform the Kaluza-Klein expansion of the fields, that is to say we separate the dependencies in the extra-dimensional coordinate $z$ and the brane coordinates $x^{\mu}$. All fields $\Phi$ are decomposed as
\begin{align}
	\Phi( x^{\mu}, z )
		&=						\sum_{n = 0}^{\infty} \Phi^{(n)}( x^{\mu} )\.\phi^{(n)}( z )
								\; ,
								\label{eq:KK}
\end{align}
wherein the $\Phi^{(n)}( x^{\mu} )$ are the (1+1)-dimensional fields and the $\phi^{(n)}( z )$ are the solutions of an eigenvalue equation. 

\begin{figure}[t]
	\begin{center} 
		\includegraphics[scale=1]{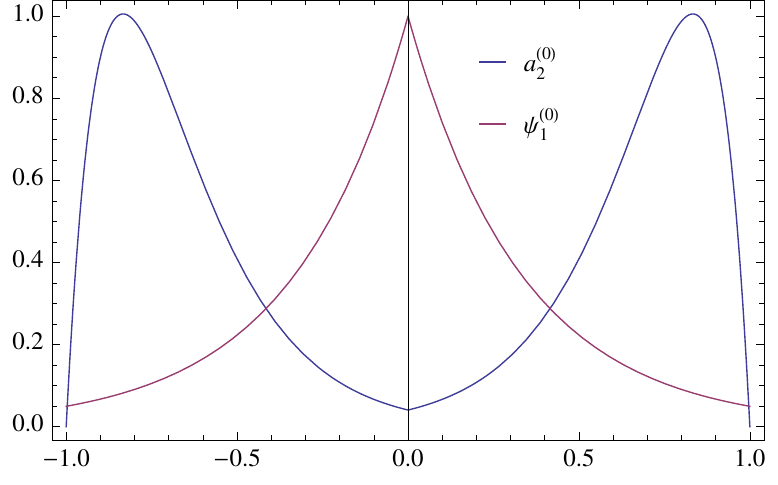}
		\caption{The lightest $a_{2}$ mode together with the lowest Fermion profile function $\psi^{(0)}_{1}$ 
				as functions of $z / L$ for $M L = 3$.\vs{-2mm}}
		\label{fig:a2-Fermion}
	\end{center}
\end{figure}

{\bf Fermionic sector:} Fermions can be easily localized following Ref.~\cite{Rubakov:1983bb} by choosing the mass functions to be $m^{}_{1}( z )= -\,m^{}_{2}( z ) = -\.M\.{\rm sign}( z )$. With $M > | \mu |$ we obtain a left-handed \footnote{The right-handed components are exponentially small.}
low-energy mode for $\Psi_1$ as well as a right-handed low-energy mode for $\Psi_2$, which read
\vs{-1mm}
\begin{align}
	\psi_{1}^{(0)}( z )
		&\simeq					\!\binom{\sqrt{M\,}\.\ee^{- M\mspace{1mu}| z |}}
								{\sqrt{M\,}\.\ee^{- 2 M L}\,\ee^{+ M\mspace{1mu}| z |}}\!
		\simeq					\gamma^{}_{1}
								\.\psi_{2}^{(0)}( z )
								\; ,
								\\[-7mm]
								\notag
\end{align}
where this expression is valid up to order $\ee^{-2 M L}$. The next-to-lightest mode has mass of order $\sqrt{M^{2} + | \mu |^{2}\,}$. We can then make the mass gap $M$ large so that it is meaningful to build a low-energy effective theory containing only the low-energy modes, \cf~Fig.~\ref{fig:a2-Fermion}.

{\bf Bosonic sector:} For Bosons the situation is more subtle, as we need to have a light Higgs, a light photon with a roughly flat wave function to satisfy charge universality, and to obtain a light axion.

The usual warp factors{\.---\.}decreasing from the brane towards the end points of the interval{\.---\.}do not fulfil all these criteria, leading to a heavy axion. Although this might not be a fundamental problem, as models with heavy axions have been studied \cite{heavy-axions}, we prefer to match the usual scenario where the axion is light.

To this aim we need a warp factor that first decreases away from the brane but then increases again close to the boundary. We checked numerically that the exact form is irrelevant as long as it decreases away from the brane as well as from the end points of the extra dimension sufficiently fast. Here{\.---\.}to have analytical solutions for the modes{\.---\.}we will consider (\cf~Fig.~\ref{fig:a2-Fermion})\footnote{As mentioned, the simple form \eqref{eq:D} of the warp factor has been chosen to get analytical results. The warp factor jumps on the brane (at $z = \pm 0$), implying that the energy-momentum tensor $T_{M N}$ also jumps there. We checked numerically that a smooth warp factor gives a $T_{M N}$, which is entirely regular. On the end points of the extra dimension, the warp factor is also singular but there nevertheless $T_{M N}$ vanishes together with its derivative (except for Fermions, for which the derivative is exponentially suppressed).}
\begin{align}
	\Delta( z )
		&=						\left( 1 - \tfrac{ | z | }{ L } \right)^{\!-1}\mspace{-1mu}\ee^{ - 2 M | z |}
								\; .
								\label{eq:D}
\end{align}

Furthermore we will work in unitary gauge, where the linearized equations for the gauge and Higgs field actually decouple.

For the Higgs, the lowest mode is symmetric\vphantom{$\frac{ 1 }{ 1 }$}, light with $m^{(0)} \simeq \sqrt{ m^{2^{^{^{}}}}_{H} + M^{2}\.8\.M L\.\ee^{- 2 M L}\,} \simeq m^{}_{H}$, and reads
\vs{-1mm}
\begin{align}
\begin{split}
	a^{(0)}( z )
		&\propto					\big( L - | z | \big)^{\!2}
								\.\ee^{( L - | z | )
								\left( \sqrt{ M^{2} - \rho_{0}^{2}\,} - M \right)}
								\\
		&\hph{\propto}				\times
								{\rm Lag}_{\mu^{}_{0}}^{\nu}
								\!\left( 2 \left( | z | - L \right)
								\sqrt{ M^{2} - \rho_{0}^{2}\,} \right)
								,
								\label{eq:a(z)}
\end{split}
\end{align}
where ${\rm Lag}_{\mu^{}_{0}}^{\nu}( \ldots )$ is the Laguerre function with $\nu = 2$, $\mu_{0} = -\,3 / 2 - M / 2 \sqrt{ M^{2} - \rho_{0}^{2}\,}$, $\rho_{0} \simeq M\.2\.\sqrt{2\.M L\,}\.\ee^{- M L}$. The other modes have masses of order $M$ and can be dropped in the low-energy field theory, so that the only low-energy mode has the expected Higgs mass.

With our particular warp factor, we are able to decouple the $A_2$ field from the photon $A_\mu$. The low-energy mode for the field $A_2$ has exactly the mass $m_{2}^{(0)} \equiv m^{}_{W}$ and reads
\vs{-2mm}
\begin{align}
	a_{2}^{(0)}( z )
		&\propto					\Delta( z )^{-1}
								\; .
								\notag
\end{align}
The next-to-lightest mode is heavy, having the mass $m_{2}^{(1)} = \sqrt{ M_{}^{2} + m_{W}^{2}\,} + \Ocal\!\left( \ee^{-2 M L} \right)$ and thus
providing again the desired mass gap.

\begin{figure}[t]
	\begin{center} 
		\includegraphics[scale=1]{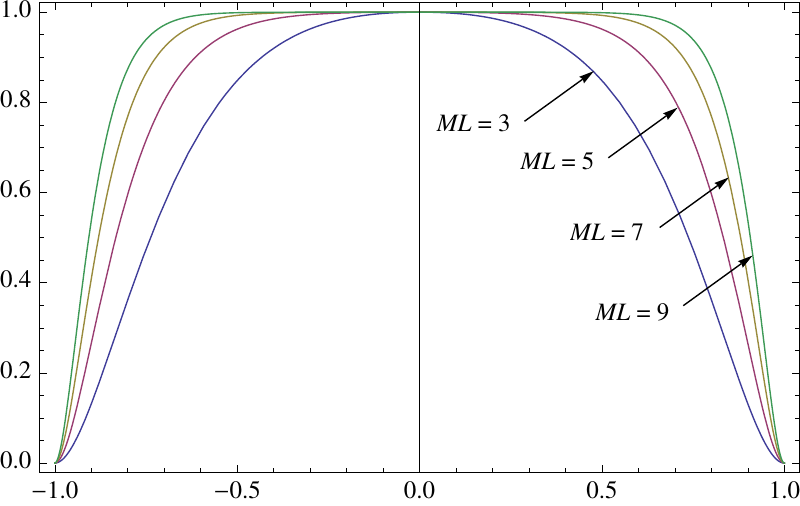}
		\caption{The lightest $a$ mode as a function of $z / L$ for various values of $M L$.\vs{-5mm}}
		\label{fig:a}
	\end{center}
\end{figure}

The equations for the photons are very similar to the Higgs case: One just has to replace $m^{}_{H}$ by $m^{}_{W}$. Hence the lightest mode is symmetric\vphantom{$\frac{ 1 }{ 1 }$}, has the mass $m^{(0)} \simeq \sqrt{ m^{2^{^{^{}}}}_{W} + M^{2}\.8\.M L\.\ee^{- 2 M L}\,} \simeq m^{}_{W}$, and the wave function given by \eqref{eq:a(z)}, \cf~Fig.~\ref{fig:a}. Also, the next-to- lightest has a mass of order $M$.\\[-4mm]

\section{Effective action}
\label{sec:Effective-action}

Choosing $M \gg m_{H}$, $m_{W}$ and $| \mu |$, we can built the effective (1+1)-dimensional action. In the initial action (\ref{eq:LH},\ref{eq:L-Fermion-3D}) we insert the Kaluza-Klein expansions \eqref{eq:KK}, truncate it to keep only the light modes, integrate over the extra-dimension, and neglect subleading terms: We get the usual (1+1)-dimensional Abelian Higgs model (\ref{eq:Lmod1}) with $\tilde{ e }^{2} \equiv e^{2}\mspace{1mu}M$, $\tilde{v}^{2} \equiv v^{2} / M$, $\tilde{\lambda} \equiv \lambda\.M$ and $\tilde{\mu} \equiv \mu$ \footnote{These identifications are exactly what we expect from dimensional analysis.}, and an additional light degree of freedom with the action
\vs{-1mm}
\begin{align}
	\!\Scal_{A_{2}}
		&\simeq					-\, \frac{ 1 }{ 2 } \int\! \d^{2}_{} x\;
								A_{2}\.\Big[
								\big( \Box + m_{W}^{2} \big) A_{2}
								- \irm\.\tilde{\rho}^{}_{\Psi}\.\bar{\Psi}\.\gamma_{5} \Psi
								\Big]
								\, ,
								\notag
\end{align}
wherein the marginal couplings to the Higgs are not shown (\cf~the previous discussion), and the $A_{2}$-Fermion coupling $\tilde{\rho}^{}_{\Psi}$ is suppressed: $\tilde{\rho}^{}_{\Psi} \simeq \tilde{e}\;2\.\sqrt{2\,}\.(M L)^{3 / 2}\.\ee^{- M L}$.

Until now, $A_{2}$ does not have the topological coupling to $A_{\mu}$. It is obtained from the Fermionic diagram \eqref{eq:qcor}, for which the Fermion loop contributes with
\begin{align}
	\put(0,-16){\includegraphics[scale=1]{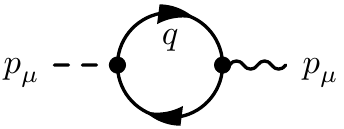}}
		&\mspace{176.5mu}=		\frac{\tilde{\rho}^{}_{\Psi} e}{2\.\pi\mspace{1mu}\tilde{\mu}}\,
								f\big( p^{2} / \tilde{\mu}^{2} \big)\.\varepsilon_{\mu\nu}\mspace{1mu}p^{\nu}
								\; .
\end{align}
In the limit of vanishing external momentum $p_{\mu}$, we find that $f\big( p^{2} / \tilde{\mu}^{2} \big) \to 1$, and finally get the effective coupling
\begin{align}
	\frac{ 1 }{ f_{a} }\.A^{}_{2}\.\varepsilon_{\mu\nu}\mspace{1mu}F^{\mu\nu}
	\q\text{with}\q
	f_{a}
		\simeq					\frac{ \sqrt{2\,}\.\pi\.\tilde{\mu}\.\ee^{M L} }{ \tilde{e}^2\.( M L )^{3 / 2}}
								\; .
								\label{eq:axion-photon-term}
\end{align}
Equation \eqref{eq:axion-photon-term} is exactly what we were after! It shows that the $A_{2}$ mode couples to the photon precisely like an axion. Moreover, the inverse axion-photon coupling constant $f_{a}$ is exponentially large, guaranteeing that phenomenological bounds are already satisfied for moderately large values of $M L$, \ie~without any fine-tuning.

Our results also rely on the applicability of perturbation theory. This can be done consistently as the warp factor is non-vanishing, leading to finite couplings everywhere in the bulk. For realistic {\sc qcd}, perturbation theory may not be applicable, however the existence of the strong-{\sc cp} problem does not depend on the strength of the coupling and lattice simulation might also been envisaged in five dimensions \cite{deForcrand:2010be}.\\[-6mm]

\section{Conclusion}
\label{sec:Conclusion}

We have seen that the occurrence of an axion could be a rather natural consequence of the presence of extra-dimensions, as it can arise from the extra-dimensional component of the gauge field. This is achieved without adding any new fields or couplings. We obtain{\.---\.}via Fermionic loop corrections{\.---\.}precisely the right axionic coupling to the topological charge with a suppressed coupling constant. The key issue is the particular selection of the boundary conditions. This ensures that the gauge anomaly{\.---\.}generically present on the class of geometries used{\.---\.}is entirely absent.\\[-6mm]

Furthermore, our results require essentially only fine-tuning concerning the warp factor, which{\.---\.}in order to get both a light axion and photon{\.---\.}should decrease sufficiently fast away from the brane as well as from the end points of the extra dimension. For this class of theories our findings are basically independent of the details of the model. For instance, the length of the extra-dimension is not really constrained, and also the dependence on the warp factor (fulfilling the mentioned requirements) is weak. A future publication will be devoted to the inclusion of gravitational dynamics to naturally obtain the warp factor.

\acknowledgments

The authors thank F.~Bezrukov, S.~Hofmann and M.~Shaposhnikov for helpful discussions. F.~K.~acknowledges the Bielefeld University for hospitality.


\end{document}